\documentclass[journal=jpcld,manuscript=article,articletitle=true,layout=twocolumn]{achemso}

\usepackage{amsmath}
\usepackage{amssymb}
\usepackage{bm}
\usepackage{latexsym}
\usepackage{graphicx}
\usepackage{epsfig,epsf,rotating}
\usepackage{indentfirst}
\usepackage{epstopdf}
\usepackage{xcolor}
\usepackage{multirow}
\usepackage{tabularx}
\usepackage{ctable}

\usepackage{upgreek}
\newcolumntype{C}{>{\centering\arraybackslash}X}
\usepackage{nicefrac}

\title{Large integration time-step in molecular dynamics simulation artificially enhances the strength of hydrophobic 
interaction\footnote{Notice:  This manuscript has been authored by UT-Battelle, LLC, under contract DE-AC05-00OR22725 with the US Department of Energy (DOE). The US government retains and the publisher, by accepting the article for publication, acknowledges that the US government retains a nonexclusive, paid-up, irrevocable, worldwide license to publish or reproduce the published form of this manuscript, or allow others to do so, for US government purposes. DOE will provide public access to these results of federally sponsored research in accordance with the DOE Public Access Plan (https://www.energy.gov/doe-public-access-plan)}}
\author{Dilipkumar N. Asthagiri}
\affiliation{Oak Ridge National Laboratory, One Bethel Valley Road, Oak Ridge, TN 37830-6012}
\email{asthagiridn@ornl.gov}

\begin{document}
\clearpage
\begin{abstract}
Molecular dynamics simulations are used to compute the potential of mean force (PMF) between two united-atom methane molecules
at several temperatures and integration time-steps.  At a fixed time-step, the contact minimum of the PMF deepens
with increasing temperature, as expected for hydrophobicity driven association. Compared with a time-step of 0.5~fs, one that 
preserves equipartition, larger time-steps alter the PMF and make the contact minimum more favorable. Thus even 
relative free energy values are sensitive to time-steps that break equipartition. Using quasichemical theory, we partition
the free energy of association into hydrophobic and hydrophilic contributions. The hydrophilic contribution opposes association and is 
insensitive to the time-step. Thus the artificial enhancement of association at larger time-steps comes entirely from the hydrophobic contribution. We explain this behavior through the temperature dependence of the internal pressure of the liquid. 
The same analysis also accounts for earlier observations of the liquid's  $p-V$ behavior under conditions that break equipartition. 
\begin{tocentry}
\center{\includegraphics{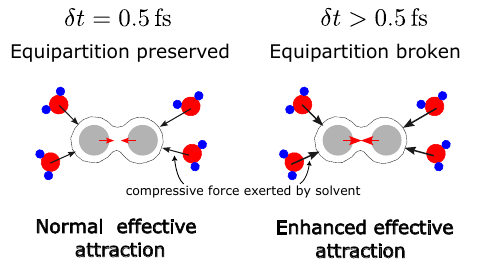}}
\end{tocentry}
\end{abstract}

\clearpage

Large-scale molecular dynamics simulations of aqueous biomolecular systems typically rely on treating the water molecule as a rigid body to enable using 
a large time-step, $\delta t$, to integrate the equations of motion. This approach rests on a long-held assumption that the fast internal vibrations in water are decoupled from its rotational and translational motion \cite{md:shake}, and hence it is permissible to use large time-steps after freezing the internal vibrations. 
Thus, $\delta t = 2$~fs has been used for decades, and more recently, $\delta t = 4$~fs has gained much popularity \cite{hopkins:hmr2015}. 

Interestingly, the initial studies on molecular dynamics simulation of liquid water, treating the water molecule as a rigid body, concluded on the basis of careful numerical experimentation that a time-step of 0.5~fs is required\cite{rahman:jcp71, stillinger:jcp74}. (These early studies predated the development of constraint algorithms that used Cartesian coordinates \cite{md:shake, andersen:rattle83, kollman:settle92}.)  Recently, we have shown that time-steps in excess of 0.5~fs for simulating the molecular dynamics of liquid water, with the water molecule treated as a rigid object, lead to a breakdown of equipartition \cite{asthagiri:jctc2024a}: for integration time-steps larger than 0.5~fs, the translational motion evolves at an effectively higher temperature than the rotational motion. This problem arises because of the separation in time-scales between translational and rotational motion in liquid water\cite{asthagiri:jctc2024a,asthagiri:jpcb2026} that necessitates the 
use of a small $\delta t$ to capture both translational and rotational relaxation with fidelity. The breakdown of equipartition impacts the hydration thermodynamics of ions and a protein in a fixed conformation \cite{asthagiri:cs2025} and the phase envelope of liquid water \cite{asthagiri:jpcb2026}. We must note that at least two research groups had also noticed the breakdown of equipartition\cite{davidchack:jcp10,abreu:jcp17} before us\cite{asthagiri:jctc2024a}. 

Time-step artifacts affect the hydration free energy of defined conformations, but does it also affect free energy \emph{changes}? In particular, do $\delta t$-artifacts affect hydrophobic association, a key driving force in biomolecular assembly\cite{kauzmann:59,tanford:62,dill:1990ww,Pratt:2002p775,chandler:nature05} and one that is intimately dependent on the properties of water itself\cite{Hummer:1996p326,lrp:jpcb98,Pratt:2002p3001}?  Here we investigate these fundamental 
questions for the potential of mean force (PMF) for the association of two methane molecules in liquid water, a system that has long anchored fundamental investigations of hydrophobicity\cite{pratt1977theory,pratt1980effects,hayment:jcp93,lrp:jpcb98,asthagiri:jcp2008,BenAmotz:jpcl15}.  The PMF is a 1D realization of more complicated multi-dimensional free energy landscapes that are of interest in biomolecular dynamics, and it is in this context that the question we pose assumes significance.  We find that the 
methane-methane PMF in water is indeed influenced by a large $\delta t$: time-step artifacts \emph{do not} cancel when considering relative values. Importantly, hydrophobic association between the methanes is amplified when liquid water is simulated with large time-steps. This finding is critical in all efforts to use computer simulations to understand the physics underlying aqueous biomolecular phenomena.
 
A detailed description of the methodology and quantification of statistical uncertainty is collected in the 
Supporting Information (SI). Here we mention only the essential details. We study the PMF of two methane molecules
in 2048 TIP4P/2005\cite{vega:tip4p}  water molecules. The methane molecules are modeled within the united-atom framework. 
The methane Lennard-Jones parameters were taken from our earlier study\cite{asthagiri:jcp2008} with one important change: 
the interaction strength was reduced by a factor of 3 to better expose the role of hydrophobicity (see below). Thus, LJ $\sigma = 3.73$~{\AA} and $\epsilon = 0.098$~kcal/mol. 
The average volume of the simulation cell at each of 283.15, 298.15, 318.15, and 338.15~K is obtained from an $NpT$ simulation with $\delta t = 0.5$~fs. In the subsequent step, the PMF between the methane atoms is obtained using stratified 
adaptive bias force sampling\cite{abf1,abf2} in the $NVT$ ensemble. For a given temperature, the volume is constrained to the average volume obtained with $\delta t = 0.5$~fs and time steps of 0.5, 2.0, and 4.0~fs are examined. We use NAMD\cite{namd,phillips2005scalable,namd:2020} to simulate the systems. 

\begin{figure*}[ht!]
\includegraphics[scale=1.1]{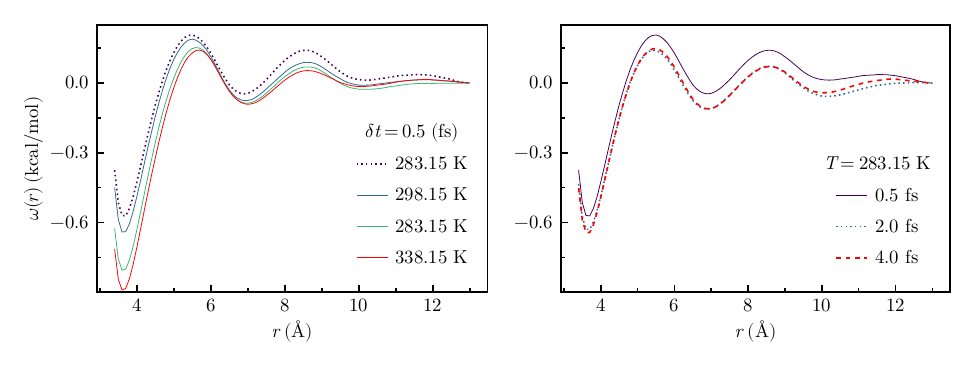}
\caption{The potential of mean force $w(r)$ along $r$, the inter-atomic separation between the united-atom methanes. The PMF is referenced relative to the value at 13~{\AA}. \underline{Left panel}: PMFs for $\delta t = 0.5$~fs but at different temperatures. \underline{Right panel}: PMFs at different time-steps but at the fixed temperature of 283.15~K. Graphs for other conditions are collected in the SI.}
\label{fg:figure1}
\end{figure*}
\textbf{PMFs}.  Figure~\ref{fg:figure1} shows the PMF between the methane united atoms. Consider first the behavior for a fixed value of $\delta t = 0.5$~fs (Fig.~\ref{fg:figure1}, left panel). As temperature increases the well depth decreases, indicating a more favorable association with temperature: this is as expected for hydrophobicity driven association and is well documented in the field. Notice also that the well-depth occurs at a separation of 3.7~{\AA}, which can be compared to the location of the minimum of the potential at $\sigma \cdot \sqrt[6]{2} \approx 4.2$~{\AA} in the absence of the solvent: the solvent pushes the solutes into contact. Turning to the right panel
shows that PMFs for $\delta t = 2.0$~fs or $\delta t = 4.0$~fs are as if the system was simulated at a higher temperature, although the thermostat temperature is 283.15~K. Thus, the discretization artifact affects the entire free energy landscape: free energy \emph{changes} are also sensitive to $\delta t$!

Fig.~\ref{fg:figure1} (and Figs.~S2 and S3) convey another point. Comparing PMFs at $\delta t = 2.0$ and $4.0$~fs, i.e.\ ``not at thermal equilibrium" results, may induce one into thinking the PMFs converge to some common answer. The problems become manifest only in comparing with results obtained under conditions ($\delta t = 0.5$~fs) that ensure thermal equilibrium.

\textbf{Hard-sphere contributions}. The PMF, $\omega(r)$, can be decomposed as
\begin{eqnarray}
\omega(r) = u(r) + \Delta \mu^{(ex)}(r) \, ,
\end{eqnarray}
where $u(r)$ is the direct interaction and $\Delta \mu^{(ex)}(r)$ is the change in the excess chemical potential of the pair of methanes as they are brought to a separation $r$ from 13~{\AA}. From the quasichemical organization of the potential distribution theorem, $\mu^{(ex)}(r)$ can be written as 
\begin{eqnarray}
\beta \mu^{(ex)}(r)  =   -\ln p_0(r) & + & \ln x_0(r) \nonumber \\
& + &  \ln \langle e^{\beta \varepsilon} | n = 0\rangle \,
\end{eqnarray}
where $\beta = 1  / k_{\rm B}T$;  $p_0(r)$ is the probability to create a pair of cavities at a separation $r$; $x_0(r)$ is the probability to evacuate the inner-shell of the same size as the cavity from around the methane atoms; and $\ln \langle e^{\beta \varepsilon} | n = 0\rangle$ is the free energy contribution from solute-water interactions ($\varepsilon$) subject to the constraint that the inner shell is bereft ($n=0$) of solvent. The first contribution on the right hand side of the equation above is precisely the primitive hydrophobic contribution to the PMF and the remaining two contributions are the hydrophilic contributions. 
Thus, the PMF can be written as 
\begin{eqnarray}
\omega(r) = u(r) + \omega_{hs}(r) + \omega_{att}(r) \, ,
\label{eq:wdecomp}
\end{eqnarray}
where $\beta \omega_{hs}(r) [= -\ln p_0(r) + \ln p_0(r = 13~\mathrm{\AA})]$ is the hard-sphere (or primitive hydrophobic) contribution and $\omega_{att}(r)$ is the contribution from hydrophilic interactions. 

We choose a hard-sphere (cavity) radius of 3.3~{\AA}, which is slightly below the peak of the methane-water oxygen pair correlation (Fig.~S5).  For this choice of radius, the conditional probability distribution $P(\varepsilon | n=0)$ admits, to an excellent approximation, a Gaussian description with mean  $\langle \varepsilon | n = 0 \rangle$ and variance $\langle \delta \varepsilon^2 | n = 0\rangle$. 
Thus, $\ln \langle e^{\beta \varepsilon} | n = 0\rangle = \beta \langle \varepsilon | n = 0 \rangle + (\beta^2/2) \langle \delta \varepsilon^2 | n = 0\rangle$. Since the hydrophilic contributions are easier to calculate, we infer the hard-sphere contribution using this, the already calculated PMF, and the known $u(r)$ (Eq.~\ref{eq:wdecomp}). 

\begin{figure}[ht!]
\includegraphics[scale=1.1]{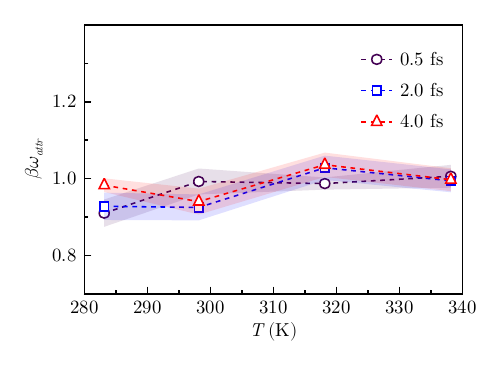}
\caption{The hydrophilic contribution (Eq.~\ref{eq:wdecomp}) as the methanes are brought from 13~{\AA} to the contact minimum. The $1\sigma$ standard error of the mean (shaded band) is obtained by block-averaging five blocks of energetic data obtained from $5 \times 5000$ configurations sampled over $5 \times 2.5\times 10^6$ steps. N.B.: The PMF (Fig.~\ref{fg:figure1}) is based on between 4x (4.0~fs) and 16x (0.5~fs) longer sampling, and hence statistical uncertainties noted above are more pronounced.}
\label{fg:figure2}
\end{figure}

Figure~\ref{fg:figure2} shows $\omega_{att}$ measured at the contact minimum. Two features stand out: (1) $\omega_{att} > 0$, i.e.\ hydrophilic interactions inhibit the association\cite{asthagiri:jcp2008,pratt1980effects}, and (2) the hydrophilic contributions are nearly flat with respect to the temperature and insensitive to $\delta t$. 

\begin{figure}[ht!]
\includegraphics[scale=1.1]{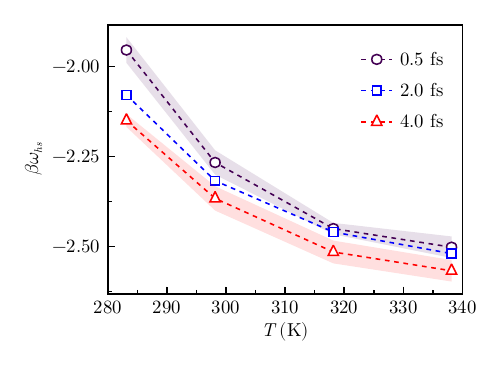}
\caption{The hydrophibic contribution (Eq.~\ref{eq:wdecomp}) to the change in the PMF as the methanes are brought from 13~{\AA} to the contact minimum. The statistical uncertainty mirrors that for $\omega_{att}$ (Fig.~\ref{fg:figure2}. The uncertainty band for $\delta t = 2.0$~fs is not shown for clarity.}
\label{fg:figure3}
\end{figure}

Figure~\ref{fg:figure3} shows  $\omega_{hs}$ calculated at the contact minimum. Together with data from Figures~\ref{fg:figure1} and~\ref{fg:figure2}, we can conclude that it is the $\delta t$-dependence of the hydrophobic contribution that is consistent with observed $\delta t$ dependence of the PMF. 

We can appreciate the implications of the errors introduced by the discretization artifact as follows. Assume the methanes are sticky spheres. Then the approximately $0.25~k_{\rm B}T$ shift in the well-depth between $\delta t = 0.5$~fs and $\delta t = 4.0$~fs at 283.15~K  will translate to a predicted increased association by a factor of $\exp(0.25) \approx 1.28$ (or 28\% greater association). Between $\delta t = 0.5$~fs and $\delta t = 2.0$~fs, the factor drops to about $\exp(0.10) \approx 1.10$ (or 10\% greater association).  Integrating $r^2 e^{-\beta \omega(r)}$ up to the first maximum can, for example, give a more refined estimate of the association. Such an integrated estimate confirms a greater association at $\delta t > 0.5$~fs than at 0.5~fs. But for reasons noted above about comparing results that are themselves not at equilibrium, 
we do not undertake such detailed calculations of association. The important take-away from Fig.~\ref{fg:figure3} and the foregoing discussion is that small changes in the PMF are expected to assume greater importance in the context of collective hydration processes as happens in biomolecular self-assembly.  

\textbf{Rationalization of $\delta t$ dependence of hydrophobicity}. The counterintuitive temperature dependence of hydrophobic interaction --- the strengthening of 
association as temperature is increased --- has drawn intense scrutiny over the decades. A key insight emerging from those studies is that hydrophobic hydration and hydrophobic association reflects the tendency of the water matrix to squeeze the nonpolar solutes out of solution \cite{Pratt:2002p3001}. This squeezing effect is evident from the observation noted earlier that the methane-methane contact minimum in the absence of the solvent is $\approx 4.2$~{\AA}, whereas in the solvent it is $\approx 3.6$~{\AA}.

We can quantify this squeezing behavior of the solvent by examining the internal pressure\cite{pratt2007special}  
\begin{eqnarray}
\left(\frac{\partial U}{\partial V}\right)_T = T\left(\frac{\alpha_p}{\kappa_T}\right) - p \, ,
\label{eq:ip}
\end{eqnarray}
where $\alpha_p \equiv -(\partial \ln \rho/\partial T)_T $ is the thermal expansion coefficient ($\rho$ is the density of neat water), $\kappa_T \equiv (\partial \ln \rho/\partial p)_T$ is the isothermal compressibility, $T$ is the temperature, and $p$ is the thermodynamic pressure. For a van~der~Waals liquid, the internal pressure $(\partial U / \partial V)_T \approx a \rho^2$, where $a$ is the van der Waals parameter and describes the effects of the attraction. For liquid hydrocarbons, this quantity gently \emph{decreases} with temperature, because $\rho$ decreases with temperature.  However, for water, $(\partial U / \partial V)_T$ \emph{increases} sharply with temperature. (See Figure~4 in Ref.~\citenum{pratt2007special}.) Eq.~\ref{eq:ip} helps understand this behavior\cite{pratt2007special}. In the temperature range of interest here, $\rho$ changes slowly, $\kappa_T$ is small and changes slowly with $T$, and
$\alpha_p$ is also small.  The ratio $\alpha_p / \kappa_T$ is a large number with a weak temperature dependence; the strong temperature dependence comes mainly from the explicit factor of $T$, a measure of the kinetic energy. The physical implication is that the squeezing effect of the liquid matrix on the solute has a dominant $T$ dependence that arises from the kick delivered to the solute by the solvent molecule's rectilinear motion. That effect is more pronounced the higher the translational kinetic energy.  When equipartition is broken for water, we find that the center-of-mass translation of the water molecule  evolves at a higher effective temperature than its rotational counterpart \cite{asthagiri:jctc2024a,asthagiri:cs2025,asthagiri:jpcb2026}. This enhanced translational kinetic energy of water has the same physical consequence as that of increasing $T$ in Eq.~\ref{eq:ip}. This is the reason behind the observed $\delta t$ dependence of the PMF and the hydrophobic contribution. This insight also concisely rationalizes the earlier finding that for the temperature range of interest and under conditions that violate equipartition ($\delta t > 0.5$~fs) the volume of the simulation box is higher (density is lower) in $NpT$ simulations\cite{asthagiri:jpcb2026} --- the enhanced translational velocity causes a greater momentum transfer to the confining walls, which is compensated by an increase in box volume to match the thermodynamic pressure. 

In conclusion, equipartition is a central tenet of equilibrium statistical mechanics, and when applied to liquid water with water molecules treated as rigid bodies, this requires that the average translational kinetic energy per water molecule equal its rotational counterpart. This requirement is violated in molecular dynamics simulations that use a large integration time-step, most notably time-steps of 2.0~fs used for decades and  a value of 4.0~fs that is now popular in large-scale biomolecular simulations. 
A key finding of this work is that $\delta t$-artifacts \emph{do not} cancel in the PMF for
the association of two methane molecules, the ``hydrogen atom" system for understanding hydrophobic association. Since the PMF is but a 1D realization of more complicated multi-dimensional free energy landscapes of interest in molecular biophysics, our work shows that sampling those multi-dimensional landscapes in simulations that violate equipartition is bound to introduce uncontrolled errors and obfuscate a clear understanding of the sought-after underlying physics. 



\section{Supporting Information}

Simulation methodology; quantification of uncertainty in the PMFs; water radial distribution around methane; calculation of hydrophilic contribution and its associated statistical uncertainty. 

\section{Acknowledgements}
Helpful discussions with Thiago Pinheiro dos Santos and Michael Paulaitis are gratefully acknowledged. This research used resources of the Oak Ridge Leadership Computing Facility at the Oak Ridge National Laboratory, which is supported by the Office of Science of the U.S. Department of Energy under Contract No. DE-AC05-00OR22725.
 
\bibliography{refs.bib}

@String{ARPC = {Ann. Rev. Phys. Chem.}}

@String{JCP  = {J. Chem. Phys.}}

@String{JCC = {J. Comp. Chem.}}

@String{JACS = {J. Am. Chem. Soc.}}

@String{JCTC  = {J. Chem. Theory Comput.}}

@String{JCoP = {J. Comput. Phys.}}

@String{JCoC = {J. Comput. Chem.}}

@String{PNAS = {Proc. Natl. Acad. Sci. USA}}

@String{CPL = {Chem. Phys. Lett.}}

@String{CR = {Chem. Rev.}}

@String{CS = {Chem. Sc.}}

@String{CoPC = {Comp. Phys. Comm.}}

@String{JPCB = {J. Phys. Chem. B}}

@String{JPCL = {J. Phys. Chem. Lett.}}

@article{pratt2007special,
  title={What is special about water as a matrix of life?},
  author={Pratt, Lawrence R and Pohorille, Andrew and Asthagiri, D},
  journal={arXiv:physics/0701282},
  year={2007}
}

@article{asthagiri:jcp2008,
author = {D. Asthagiri and S. Merchant and L. R. Pratt}, 
journal = JCP, 
title = {Role of 
Attractive Methane-Water Interactions in the Potential of Mean Force Between Methane Molecules in Water},
pages = {244512},
volume = {128},
year = {2008}
}

@article{Hummer:1996p326,
author = {G Hummer and S Garde and A E Garcia and A Pohorille and L R Pratt}, 
journal = PNAS,
title = {An Information Theory Model of Hydrophobic Interactions},
pages = {8951-8955},
volume = {93},
year = {1996}
}

@article{Pratt:2002p3001,
author = {L. R. Pratt}, 
journal = ARPC,
title = {Molecular Theory of Hydrophobic Effects: ``{She} Is Too Mean To Have Her Name Repeated"},
pages = {409--436},
volume = {53},
year = {2002}
}

@article{Pratt:2002p775,
author = {L.~R. Pratt and A. Pohorille}, 
journal = CR,
title = {Hydrophobic Effects and Modeling of Biophysical Aqueous Solution Interfaces},
pages = {2671-2692},
volume = {102},
year = {2002}
}

@article{lrp:jpcb98,
Journal = JPCB,
Year = {1998},
Title = {Hydrophobic Effects on a Molecular Scale},
Pages = {10469-10482},
Author = {Hummer, G. and Garde, S. and Garcia, A. E. and Paulaitis,
                  M. E. and Pratt, L. R.},
Volume = {102}}

@Article{namd,
  author = 	 {L. Kale and R. Skeel and M. Bhandarkar and
R. Brunner and A. Gursoy and N. Krawetz and J. Phillips and
A. Shinozaki and K. Varadarajan and K. Schulten},
  title   = {NAMD2: Greater Scalability for Parallel Molecular
Dynamics}, 
  journal = 	 JCoP,
  year = 	 {1999},
  volume =	 {151},
  pages =	 {283}
}

@article{phillips2005scalable,
  title={Scalable molecular dynamics with NAMD},
  author={Phillips, James C and Braun, Rosemary and Wang, Wei and Gumbart, James and Tajkhorshid, Emad and Villa, Elizabeth and Chipot, Christophe and Skeel, Robert D and Kale, Laxmikant and Schulten, Klaus},
  journal=JCC,
  volume={26},
  number={16},
  pages={1781--1802},
  year={2005},
  publisher={Wiley Online Library}
}

@article{namd:2020,
author = {Phillips,James C.  and Hardy,David J.  and Maia,Julio D. C.  and Stone,John E.  and Ribeiro,João V.  and Bernardi,Rafael C.  and Buch,Ronak  and Fiorin,Giacomo  and Hénin,Jérôme  and Jiang,Wei  and \emph{et al.}},
title = {{Scalable Molecular Dynamics on CPU and GPU Architectures with NAMD}},
journal = JCP,
volume = {153},
number = {4},
pages = {044130},
year = {2020},
doi = {10.1063/5.0014475}}

@article{pratt1980effects,
  title={Effects of Solute--Solvent Attractive Forces on Hydrophobic
                  Correlations}, 
  author={Pratt, Lawrence R and Chandler, David},
  journal={J. Chem. Phys.},
  volume={73},
  pages={3434--3441},
  year={1980},
  publisher={AIP}
}

@article{pratt1977theory,
  title={Theory of the hydrophobic effect},
  author={Pratt, Lawrence R and Chandler, David},
  journal={J. Chem. Phys.},
  volume={67},
  number={8},
  pages={3683--3704},
  year={1977},
  publisher={American Institute of Physics}
}

@Article{chandler:nature05,
  author = 	 {D. Chandler},
  title = 	 {Interfaces and the driving force of hydrophobic assembly},
  journal = 	 {Nature},
  year = 	 {2005},
  volume =       {437},
  pages =	 {640-647}
}

@Article{md:shake,
  author = 	 {J. P. Ryckaert and G. Ciccotti and H. J. C. Berendsen},
  title = 	 {Numerical Integration of the {Cartesian} Equations
                  of Motion of a System with Constraints: Molecular
                  Dynamics of n-alkanes},
  journal = 	 JCoP,
  year = 	 {1977},
  volume =	 {23},
  pages =	 {327-341}
}

@Article{merchant:jcp11a,
  author = 	 {S. Merchant and J. K. Shah and D. Asthagiri},
  title = 	 {Water Coordination Structures and the Excess Free
                  Energy of the Liquid},
  journal = 	 JCP,
  year = 	 {2010},
  volume =	 {134},
  pages =	 {124514}
}

@Article{tanford:62,
  author = 	 {C. Tanford},
  title = 	 {Contribution of Hydrophobic Interactions to the
                  Stability of the Globular Conformation of Proteins},
  journal = 	 JACS,
  year = 	 {1962},
  volume =	 {84},
  pages =	 {4240--4247}
}

@Article{kauzmann:59,
  author = 	 {W. Kauzmann},
  title = 	 {Some Factors in the Interpretation of Protein Denaturation},
  journal = 	 {Adv. Prot. Chem.},
  year = 	 {1959},
  volume =	 {14},
  pages =	 {1--63}
}

@article{dill:1990ww,
author = {Dill, K. A.},
title = {{Dominant Forces in Protein Folding}},
journal = {Biochem.},
year = {1990},
volume = {29},
pages = {7133--7155}
}

@Article{vega:tip4p,
  author = 	 {J. L. F. Abascal and C. Vega},
  title = 	 {A General Purpose Model for the Condensed Phases of Water: {TIP4P/2005}},
  journal = 	 JCP,
  year = 	 {2005},
  volume =	 {123},
  pages =	 {234505}
}

@Article{abf1,
  author = 	 {E. Darve and D. Rodriguez-G{\'o}mez and A. Pohorille},
  title = 	 {Adaptive biasing force method for scalar and vector
                  free energy calculations},
  journal = 	  JCP,
  year = 	 {2008},
  volume =	 {128},
  pages =	 {144120}
}

@Article{abf2,
  author = 	 {J. H{\'e}nin and G. Fiorin and C. Chipot and M. L. Klein},
  title = 	 {Exploring multidimensional free energy landscapes
                  using time-dependent biases on collective variables},
  journal = 	 JCTC,
  year = 	 {2010},
  volume =	 {6},
  pages =	 {35--47}
}

@article{BenAmotz:jpcl15,
author = {D. Ben-Amotz},
title = {{Hydrophobic Ambivalence: Teetering On The Edge Of Randomness}},
journal = JPCL,
year = {2015},
volume = {6},
pages = {1696--1701}
}

@article{stillinger:jcp74,
	title = {Improved Simulation of Liquid Water by Molecular Dynamics},
	volume = {60},
	journal = JCP,
	author = {Stillinger, Frank H. and Rahman, Aneesur},
	year = {1974},
	pages = {1545--1557}
}

@article{rahman:jcp71,
	title = {Molecular {Dynamics} {Study} of {Liquid} {Water}},
	volume = {55},
	journal = JCP,
	author = {Rahman, Aneesur and Stillinger, Frank H},
	year = {1971},
	pages = {3336--3359},
}

@article{andersen:rattle83,
	title = {Rattle: {A} Velocity Version of the {Shake} Algorithm
                  for  Molecular Dynamics Calculations},   
	volume = {52},
	journal = JCoP,
	author = {Andersen, Hans C.},
	year = {1983},
	pages = {24--34}
}

@article{kollman:settle92,
	title = {Settle: {An} Analytical Version of the {SHAKE} and {RATTLE} Algorithm for Rigid Water Models},
	volume = {13},
	journal = JCoC,
	author = {Miyamoto, Shuichi and Kollman, Peter A},
	year = {1992},
	pages = {952--962}
}

@Article{svr:jcp07,
  author = 	 {G. Bussi and D. Donadio and M. Parrinello},
  title = 	 {Canonical Sampling Through Velocity Rescaling},
  journal = 	 JCP,
  year = 	 {2007},
  volume =	 {126},
  pages =	 {014101}
}

@Article{davidchack:jcp10,
  author = 	 {R. L. Davidchack},
  title = 	 {Discretization Errors in Molecular Dynamics
                  Simulations with Deterministic and Stochastic
                  Thermostats}, 
  journal = 	 JCoP,
  year = 	 {2010},
  volume =	 {229},
  pages =	 {9323-9346}
}

@Article{abreu:jcp17,
  author = 	 {Ana J. Silveira and C. R. A. Abreu},
  title = 	 {Molecular Dynamics with Rigid Bodies: Alternative
                  Formulation and Assessment of its Limitations when
                  Employed to Simulate Liquid Water},
  journal = 	 JCP,
  year = 	 {2017},
  volume =	 {147},
  pages =	 {124104}
}

@article{asthagiri:jctc2024a,
	title = {{MD} Simulation of Water using a Rigid Body
                  Description Requires a Small Time Step to Ensure
                  Equipartition}, 
	volume = {20},
	issn = {1549-9618},
	doi = {10.1021/acs.jctc.3c01153},
	number = {1},
	journal = JCTC,
	author = {Asthagiri, Dilipkumar N. and Beck, Thomas L.},
	month = jan,
	year = {2024},
	pages = {368--374},
}

@article{asthagiri:cs2025,
	title = {Consequences of the Failure of Equipartition for the
                  $p-V$ Behavior of Liquid Water and the Hydration
                  Free Energy Components of Small Protein}, 
	volume = {16},
	journal = CS,
	author = {Asthagiri, Dilipkumar N. and Valiya~Parambathu,
                  Arjun. and Beck, Thomas L.},
	year = {2025},
	pages = {7503-7512}
}

@article{asthagiri:jpcb2026,
	title = {Equipartition and the {Temperature} of {Maximum} {Density} of {TIP4P}/2005 {Water}},
	volume = {130},
	journal = JPCB, 
	author = {Asthagiri, Dilipkumar N. and Pinheiro Dos Santos, Thiago and Beck, Thomas L.},
	year = {2026},
	pages = {862-867}
}

@article{chow:mcbarostat1995,
	title = {Isothermal-isobaric Molecular Dynamics Simulations
                  with {Monte} {Carlo} Volume Sampling}, 
	journal = CoPC,
	author = {Chow, Kim-Hung and Ferguson, David M},
	year = {1995},
	volume = {91},
	pages = {282-289}
}

@article{aqvist:mcbarostat2004,
	title = {Molecular Dynamics Simulations of Water and
                  Biomolecules with a {Monte} {Carlo} Constant
                  Pressure Algorithm}, 
	volume = {384},
	doi = {10.1016/j.cplett.2003.12.039},
	journal = CPL,
	author = {{\AA}qvist, Johan and Wennerstr{\"o}m, Petra and Nervall, Martin and Bjelic, Sinisa and Brandsdal, B. O.},
	year = {2004},
	pages = {288--294}
}

@article{hopkins:hmr2015,
	title = {Long-{Time}-{Step} {Molecular} {Dynamics} through {Hydrogen} {Mass} {Repartitioning}},
	volume = {11},
	doi = {10.1021/ct5010406},
	number = {4},
	journal = JCTC,
	author = {Hopkins, Chad W. and Le Grand, Scott and Walker, Ross C. and Roitberg, Adrian E.},
	year = {2015},
	pages = {1864--1874}
}

@article{packmol,
title = {Packmol: A package for building initial configurations for
                  molecular dynamics simulations},
author = {Mart\'inez, L. and Andrade, R. and E. G. Birgin and
                  J. M. Mart\'inez},
journal = JCC,
volume = {30},
year = {2009},
pages = {2157-2164}}

@article{hayment:jcp93,
	title = {Free energy, entropy, and internal energy of hydrophobic interactions: {Computer} simulations},
	volume = {98},
	journal = JCP, 
	author = {Smith, David E. and Haymet, A. D. J.},
	year = {1993},
	pages = {6445--6454}
}

\end{document}


\newpage

\tableofcontents

\clearpage

 \section{Methodology} 
 
 We used NAMD\cite{namd,phillips2005scalable,namd:2020} for all simulations. Water was modeled using the TIP4P/2005 potential \cite{vega:tip4p}, with the geometry of water maintained using the default SETTLE\cite{kollman:settle92} algorithm.  The simulation system comprised 2048 water molecules. Methane was modeled as a united atom with Lennard-Jones $\varepsilon = 0.098$~kcal/mol, a third of the value used  in our previous work \cite{asthagiri:jcp2008}, and $\sigma = 3.73$~{\AA}. Previously\cite{asthagiri:jcp2008} we had shown that attractive interactions counter the hydrophobicity-driven association between the methane molecules. So to better isolate the  role of time-step, $\delta t$, on hydrophobic hydration, we dialed down the attraction.
 
 The Lennard-Jones interactions and real-space electrostatic interactions were cut at 10~{\AA}. Long-range, analytical LJ corrections were applied.  Long-range electrostatic interactions were obtained using Particle Mesh Ewald with a grid spacing of 1~{\AA} and 
the relative Ewald energy tolerance at the real-space cutoff is $10^{-7}$, which is an order of magnitude tighter than the default in NAMD.
 
We studied methane-methane association at temperatures of 283.15, 298.15, 318.15, and 338.15~K, respectively. 

\subsection{Initial $NpT$ simulation with $\delta t = 0.5$~fs}

We first obtained the equilibrium volumes at the temperature of interest by simulating a pair of methanes in water. Initially, the methane atoms were placed along the diagonal a distance 6~{\AA} apart in a box with the density of water ($\approx 0.97$ gm/cc) somewhat lower than the equilibrium density at the temperature of interest. The system was prepared using Packmol\cite{packmol}. After 1000 steps of energy minimization, we simulated the system under $NpT$ conditions for $3\times 10^6$ steps with $\delta t = 0.5$~fs, i.e.\ for a simulation time of 1.5~ns. The canonical stochastic velocity rescaling thermostat (CSVR)\cite{svr:jcp07} was used for temperature control and the Monte Carlo barostat \cite{chow:mcbarostat1995,aqvist:mcbarostat2004} was used to target a pressure of 1 bar.  Within the barostat, volume moves were sampled every 80 steps. Simulation details were logged every 500 steps (0.250~ps). Conservatively, the volumes and temperature relax well within 30~ps, but we average data only after 500~ps of simulation. 

\setlength{\tabcolsep}{12pt}
\ctable[
	mincapwidth=\textwidth,
	caption={Average volume and length of the cubic simulation cell. The volume was averaged using data collected in the last 1~ns of a 1.5~ns simulation. The $1\sigma$ standard error of the mean volume is noted. The time-step is 0.5~fs.},
	label=tb:volumes,
	pos=h,
	captionskip=-1.ex
	]	
{c c c}
{}
{
\FL
$T$ (K) &  V ({\AA}$^3$) &  $L$ ({\AA})  \ML
283.15  &  $61397\pm 22$ & 39.45   \NN
298.15  &  $61583\pm 16$ & 39.49   \NN
318.15  &   $62063\pm 21$ & 39.59   \NN
338.15  &  $62703\pm 16$  & 39.73   \LL
}

The average volume (Table~\ref{tb:volumes}) was used unchanged in the next phase where we study the impact of $\delta t$ on the potential of mean force (PMF) between the methanes. Note that as shown previously\cite{asthagiri:cs2025,asthagiri:jpcb2026} time-step artifacts also affect the $p-V$ behavior. Specifically, the volume for $\delta t > 0.5$~fs is found to be larger than the volume with $\delta t = 0.5$~fs. Those $p-V$ changes can obfuscate the role of $\delta t$ on the PMF, and hence we chose to keep the volumes constant at the value obtained with the time-step that guarantees equipartition\cite{asthagiri:jctc2024a,asthagiri:jpcb2026}.

 \subsection{PMF with stratified adaptive bias-force sampling} 
 
 In the next phase, for each temperature noted in Table~\ref{tb:volumes}, we prepared an
 ensemble of eight (8) systems $\{\mathrm{Ensemble0}, \ldots , \mathrm{Ensemble7}\}$ with the methanes separated by 4.5, 5.5, 6.5, 7.5, 8.5, 9.5, 10.5, and 11.5~{\AA}, respectively. As above, the methane molecules were initially placed along the diagonal and the system energy minimized, while keeping the methanes fixed in space. The volume of the system was based on Table~\ref{tb:volumes}.  

We then perform adaptive bias force (ABF) sampling for each member of the ensemble. The separation between the methane atoms, $r$, served as the order parameter in the ABF sampling. The order parameter is restrained to lie within the domain indicated in Table~\ref{tb:stratified} below by means of a harmonic confining potential applied at either edge of the domain. This so-called harmonic wall has a force constant of 50~kcal/mol-{\AA}$^2$.
\ctable[
	mincapwidth=\textwidth,
	caption={Stratified adaptive bias force (ABF) sampling. The domain (3.2~{\AA}, 13.0~{\AA}) is split into overlapping strata. For each member of the ensemble, the domain for the order parameter is indicated. All lengths are in units of {\AA}.},
	label=tb:stratified,
	pos=h,
	captionskip=-1.ex
	]	
{c c}
{}
{
\FL
Ensemble \# &  $r \in$ domain \ML
0                   & (3.2, 5.5) \NN
1                   & (4.5, 6.5) \NN
2                   & (5.5, 7.5) \NN
3                   & (6.5, 8.5) \NN
4                   & (7.5, 9.5) \NN
5                   & (8.5, 10.5) \NN
6                   & (9.5, 11.5) \NN
7                   & (10.5, 13.0) \LL
}

In ABF sampling, the domain is split into bins of 0.1~{\AA} and the count of the order parameter accumulated as the simulation progresses. From these counts a biasing force is constructed to 
nudge the simulation towards a uniform (diffusive) sampling of $r$ in its respective domain. 
Under the condition of uniform sampling, the underlying potential of mean force, which is the quantity of interest, is then recovered from the biasing forces that serve to cancel the force that arises from the PMF. Drawing a parallel with 
expanded ensemble sampling\cite{merchant:jcp11a}, monitoring the (biased) probability distribution of $r$ is a convenient way to monitor how well the biasing forces converge. In this study, we first do a burn-in phase of $2.5\times 10^6$ steps after which we start monitoring the sampling. On this basis, and to be conservative, we simulated the ensembles for the times noted in Table~\ref{tb:samplingt}. 
\ctable[
	mincapwidth=\textwidth,
	caption={Sampling length $N$ (steps) after an initial burn-in phase of $2.5\times 10^6$ steps. Note that each member of the ensemble was simulated for this length of time.  The sampling time, $N_t$, is obtained from the number of steps and the corresponding time-step.},
	label=tb:samplingt,
	pos=h,
	captionskip=-1.ex
	]	
{c c c}
{}
{
\FL
$\delta t$ (fs)     &  N / $10^6$ (steps)   &   $N_t$ (ns) \ML
0.5                     &  80              &  40   \NN
2.0                     &  20             &  40  \NN
4.0                     &  10             &  40  \LL
}

Thus, for example, the reconstructed methane-methane PMF at 283.15 K using a time step of 0.5~fs was based on a total sampling of  $8\times 40 = 320$~ns or $640\times 10^6$ steps. As the table above shows, for consistency all the simulations were propagated for
the same length of time. 

\begin{figure}[ht!]
\includegraphics[scale=1.0]{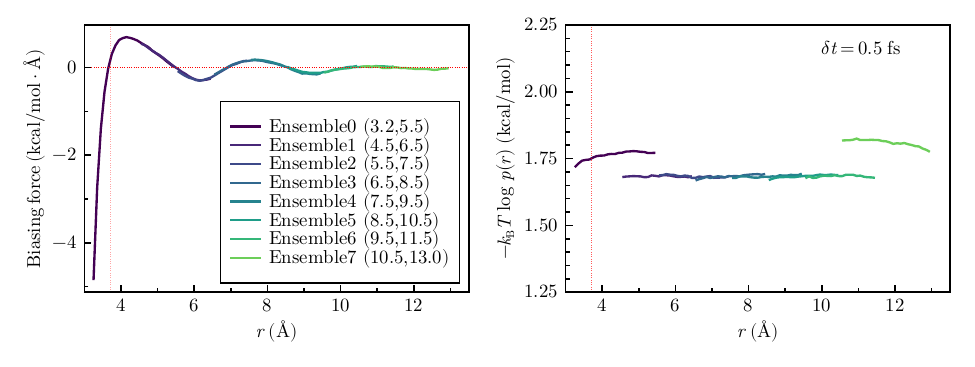} 
\includegraphics[scale=1.0]{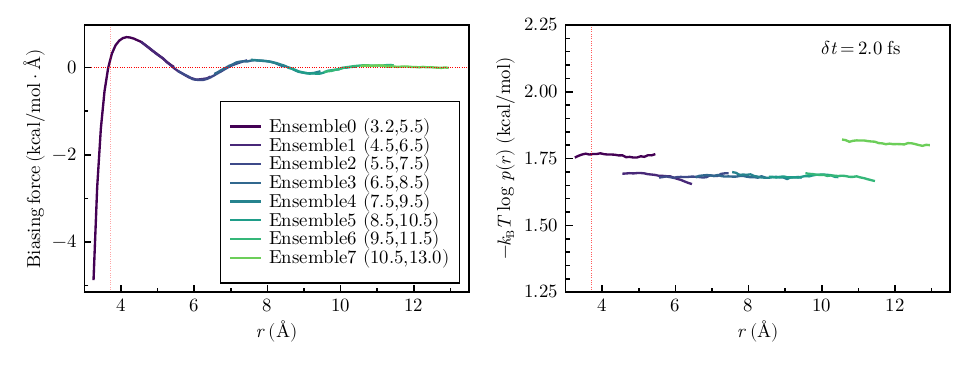} 
\includegraphics[scale=1.0]{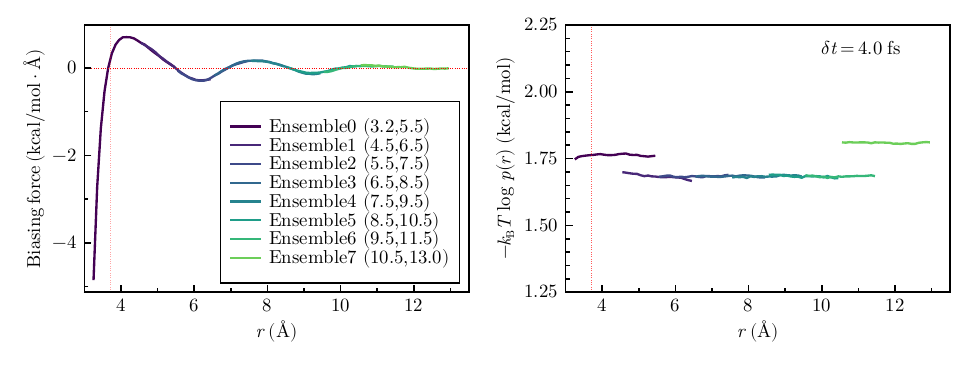} 
\caption{Figures showing overlap of biasing forces (left panel) and the corresponding biased probability distribution of $r$ (right panel). The simulation temperature is 283.15~K. For higher temperatures, the sampling is as good as if not better than what is shown. The intersection of the dotted red lines identifies the location of the minimum of the PMF at 3.7~{\AA}. Top panel: $\delta t = 0.5$~fs; middle panel: $\delta t = 2.0$~fs; bottom panel: $\delta t = 4.0$~fs.}
\label{fg:sampling}
\end{figure}

Figure~\ref{fg:sampling} documents the quality of sampling at 283.15~K for the three different
time-steps used in this study. Note that in reconstructing the PMF, the forces in the overlapping strata are averaged to get the average biasing force for $r\in (3.2, 13.0)$~{\AA}. 
Note that $p(r)$ is slightly shifted for the first and last ensembles. This is simply a consequence of a slight shift in the normalization: as the ABF procedure progresses, counts are accumulated at defined intervals, and occasionally, the order parameter $r$ can slip out of the domain to which it is restrained. Those cases are not counted and that leads to slightly varying normalizations. This shift is inconsequential in determining the forces and hence the PMF. 

\begin{figure}[ht!]
\includegraphics[scale=1.0]{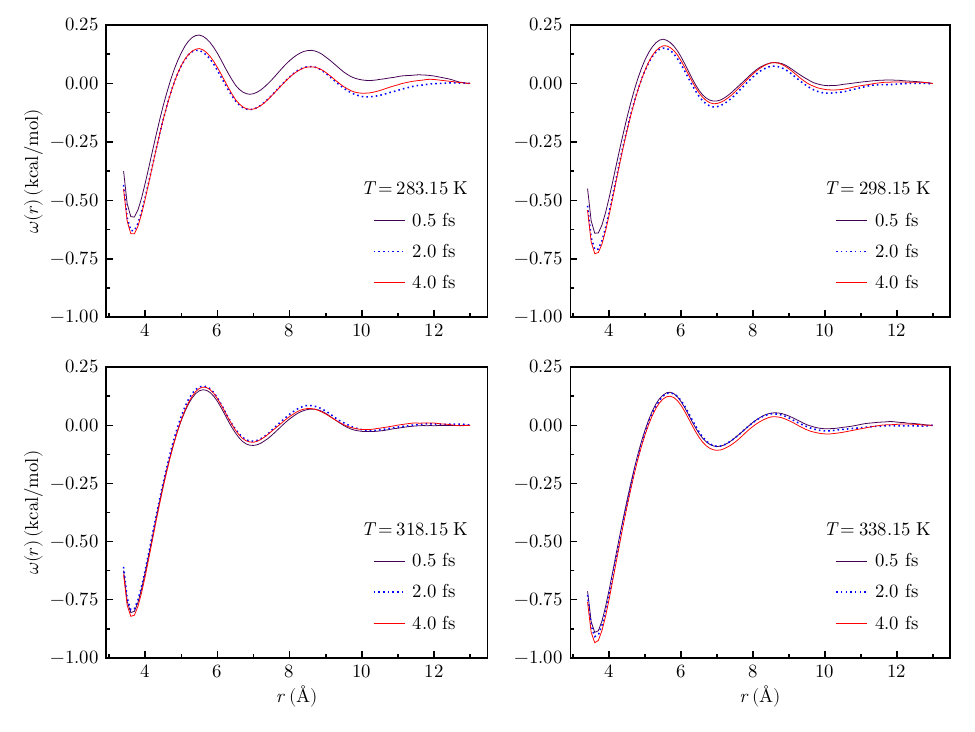} 
\caption{PMFs for all cases studied in this work.} 
\label{fg:pmfs}
\end{figure}

Figure~\ref{fg:sampling} shows that the biased probability distribution is flat to within 0.05~kcal/mol (or $\approx 0.1$~k$_{\rm B}$T). The well-depth in the PMF on the other hand is about 10x larger, being approximately 0.5~kcal/mol (Fig.~\ref{fg:pmfs}), emphasizing 
that the differences in PMFs across $\delta t$ are statistically meaningful. 

\subsection{Quantifying uncertainty on the PMF} 

To estimate the uncertainty on the PMF, we need to estimate the uncertainty on the biasing forces. In the ABF procedure, the counts of the order parameter $r$ continue to be collected as the simulation progresses. This means that the biasing forces are correlated along the length of the sampling. Visual examination of the biasing forces after each block of $2.5\times 10^6$ steps (not shown)
indicates that the biasing forces tend to stabilize and plateau. We could, in principle, look at the entire history of the biasing force in each bin and from a time-correlation analysis obtain a mean and standard error of the mean. However, this procedure proves to be demanding. 

\begin{figure}[ht!]
\includegraphics[scale=1.5]{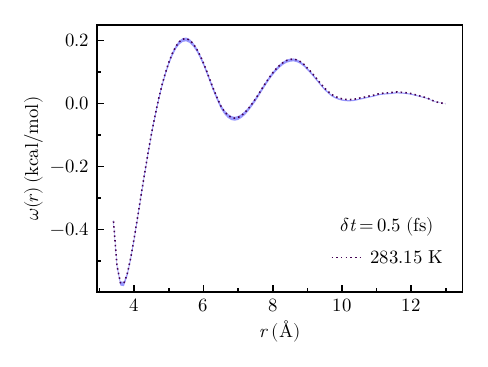} 
\caption{The PMF for methane association at 283.15~K with the $2\sigma$ standard error of the mean value shown as a shaded band obtained over 5 simulations starting from the end-point of the original sampling (Table~\ref{tb:samplingt}).} 
\label{fg:pmferr1}
\end{figure}

Instead, focusing on the case of $T=283.15$~K and $\delta t = 0.5$~fs, we launched five separate simulations of length $2.5\times 10^6$ steps from the end point of our original sampling (Table~\ref{tb:samplingt}). For these five simulations, we used the same starting coordinate but sampled fresh velocities to initiate the MD simulation. From the five PMFs obtained at the end of this procedure, we calculated the mean PMF and the standard error of the mean. Figure~\ref{fg:pmferr1} shows the original PMF value and the $2\sigma$ standard error of the mean value. 
The small spread of the PMF values suggests that our original estimate of the PMF is credible, and importantly, the deviations between PMFs (for different time-steps) are meaningful. 

\begin{figure}[ht!]
\includegraphics[scale=1.0]{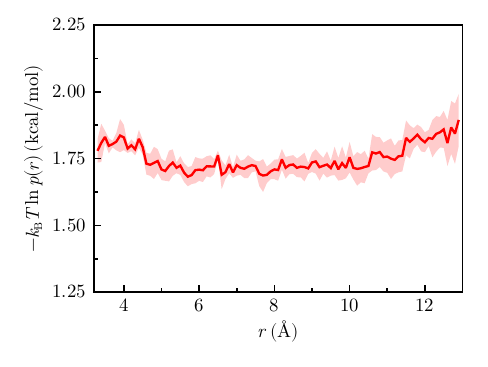} 
\caption{Mean and standard error of the distribution of $r$ using biasing forces that are fixed at the value at the end of original sampling (Table~\ref{tb:samplingt}). The $2\sigma$ standard error of the mean value is shown by the shaded background.}
\label{fg:prsampling}
\end{figure}

Another check we performed was to check the uniformity of the sampling of $r$ using the converged biasing forces that are frozen at the value obtained at the end of the original sampling (Table~\ref{tb:samplingt}). To this end, for $T=283.15$~K and $\delta t = 0.5$~fs, we launched a fresh simulation of $2.5\times 10^6$ steps archiving coordinates every 500 steps. From the archived trajectories, we calculate the probability distribution of $r$ in each member of the ensemble and then combine the result. We repeated this process four more times to obtain a mean and standard error of the mean (over the 5 independent runs). Figure~\ref{fg:prsampling} shows that the short length of simulation notwithstanding, the predicted $p(r)$ distribution is flat to within 0.25 kcal/mol ($\approx 0.4$~k$_{\rm B}$T). Note that the original sampling (Table~\ref{tb:samplingt}) was 16 times longer. With that length of the sampling, we expect the $p(r)$ distribution to be essentially flat with a significantly small uncertainty. 

\section{Methane-water oxygen pair correlation}

\begin{figure}[ht!]
\includegraphics[scale=1.0]{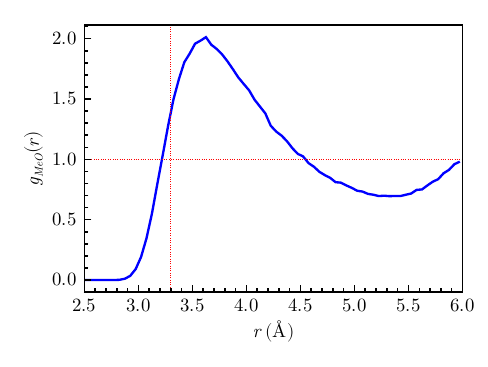} 
\caption{The methane-water oxygen radial distribution function. Only the region around the first shell is shown for clarity. The vertical red dotted line is chosen as the hard-sphere radius.}
\end{figure}

\section{Assessing the contribution of hydrophilic hydration}

The PMF, $\omega(r)$,between the methane particles can written as 
\begin{eqnarray}
\omega(r) = u(r) + \omega_{hs}(r) + \omega_{att}(r) \, 
\end{eqnarray}
where $u(r)$ is the direct interaction energy, $\omega_{hs}(r)$ is the hard-sphere (or primitive hydrophobic) contribution to hydration, and $\omega_{att}(r)$ is the attractive contribution. The latter two contributions are, of course, solvent mediated and of first interest. 
 $\omega_{hs}(r)$ depends on the specification of the hard-sphere radius, and following earlier work\cite{asthagiri:jcp2008} we use a radius of 3.3 {\AA}. We reference the PMF relative to the value at 13~{\AA}, and we are interested in the $\delta t$-dependence of the change  in $\omega_{att}$ from 13~{\AA} to the minimum of the PMF. 
 
From the quasichemical organization of the potential distribution theorem \cite{asthagiri:jcp2008}, we can write the excess chemical potential of a methane atom in water as 
\begin{eqnarray}
\beta \mu^{(\rm ex)}_{Me} & = & -\ln p_0 + \ln x_0 + \ln \langle e^{\beta \varepsilon} | n = 0\rangle  \\ 
  & \approx & -\ln p_0 + \ln x_0 + \beta \langle \varepsilon | n = 0 \rangle + \frac{\beta^2}{2} \langle \delta \varepsilon^2 | n = 0\rangle \, 
\end{eqnarray}
where $\beta = 1/k_{\rm B}T$, $p_0$ is the probability of opening a cavity of a defined size in the liquid, and $x_0$ is the probability of observing an empty inner shell, of the same size as the cavity, around the solute in the liquid; $\langle \varepsilon | n = 0 \rangle$ is the solute-solvent binding energy conditioned on the inner shell being empty, and $\langle \delta \varepsilon^2 | n = 0\rangle$ is the variance of the said binding energy distribution. For a cavity/inner-shell of radius 3.3~{\AA}. We find that the Gaussian model (Eq.~3) is an excellent approximation to $ \ln \langle e^{\beta \varepsilon} | n = 0\rangle$. In the above equations, the quantities besides $\ln p_0$ together define the hydrophilic contribution to hydration. These notions carry over unchanged in estimating the hydrophilic contribution to the PMF, for now instead of a single solute, we treat the pair of solutes as the solute. 

To estimate $\Delta \omega_{att}$, we simulated pairs of methanes fixed in space at either the contact minimum or at 13~{\AA} for a total length of $5 \times 2.5\times 10^6$ steps archiving frames every 500 steps. Each block of 5000 frames provides an estimate of 
the hydrophilic contribution. The values from the five blocks are combined to get a mean and the standard error of the mean. Since, effectively we are obtaining energetics over only 5000 frames from $2.5\times 10^6$ steps, we can see that the length of the sampling for the PMF itself is between 4x (for 4.0~fs) and 16x (for 0.5~fs) longer.

\clearpage

\bibliography{refs.bib}